\documentclass[11pt,a4paper]{emulateapj}

\usepackage{natbib}
\usepackage{latexsym}
\usepackage{url}
\usepackage[utf8]{inputenc}
\usepackage{fancyref}
\usepackage{hyperref}
\hypersetup{colorlinks,breaklinks,
            urlcolor=[rgb]{0.2,0.2,0.2},  
            linkcolor=[rgb]{0,0,0},  
            citecolor=[rgb]{0,0.4,1.0}}  

\newcommand{\teff}{${T}_{\mathrm{eff}}$}
\newcommand{\logg}{$\log{g}$}
\newcommand{\msun}{${M}_{\odot}$}
\newcommand{\mstar}{${M}_{\star}$}
\newcommand{\muhz}{$\mu$Hz}

\shorttitle{Precision asteroseismology of GD 1212 using a two-wheel-controlled {\em Kepler} spacecraft}
\shortauthors{Hermes et al.}

\begin{document}

\title{PRECISION ASTEROSEISMOLOGY OF THE PULSATING WHITE DWARF GD 1212 USING A TWO-WHEEL-CONTROLLED {\em KEPLER} SPACECRAFT}

\author{
J.~J.~Hermes\altaffilmark{1},
S.~Charpinet\altaffilmark{2,3},
Thomas~Barclay\altaffilmark{4,5},
E.~Pak\v stien\.e\altaffilmark{6},
Fergal~Mullally\altaffilmark{4,7},
Steven~D.~Kawaler\altaffilmark{8},
S.~Bloemen\altaffilmark{9},
Barbara~G.~Castanheira\altaffilmark{10},
D.~E.~Winget\altaffilmark{10},
M.~H.~Montgomery\altaffilmark{10},
V.~Van~Grootel\altaffilmark{11,12},
Daniel~Huber\altaffilmark{4,7},
Martin~Still\altaffilmark{4,5},
Steve~B.~Howell\altaffilmark{4},
Douglas~A.~Caldwell\altaffilmark{4,7},
Michael~R.~Haas\altaffilmark{4},
and~Stephen~T.~Bryson\altaffilmark{4}
}

\altaffiltext{1}{Department of Physics, University of Warwick, Coventry\,-\,CV4~7AL, United Kingdom}
\altaffiltext{2}{Universit\'e de Toulouse, UPS-OMP, IRAP, Toulouse, France}
\altaffiltext{3}{CNRS, IRAP, 14 Av. E. Belin, 31400 Toulouse, France}
\altaffiltext{4}{NASA Ames Research Center, Moffett Field, CA 94035, USA}
\altaffiltext{5}{Bay Area Environmental Research Institute, 596 1st Street West, Sonoma, CA 95476, USA}
\altaffiltext{6}{Institute of Theoretical Physics and Astronomy, Vilnius University, Gostauto 12, Vilnius LT-01108, Lithuania}
\altaffiltext{7}{SETI Institute, 189 Bernardo Ave, Mountain View, CA 94043, USA}
\altaffiltext{8}{Department of Physics and Astronomy, Iowa State University, Ames, IA 50011, USA}
\altaffiltext{9}{Department of Astrophysics, IMAPP, Radboud University Nijmegen, PO Box 9010, NL-6500 GL Nijmegen, The Netherlands}
\altaffiltext{10}{Department of Astronomy and McDonald Observatory, University of Texas at Austin, Austin, TX\,78712, USA}
\altaffiltext{11}{Institut d'Astrophysique et de G\' eophysique, Universit\' e de Li\` ege, 17 All\'ee du 6 Ao\^ ut, B-4000 Li\` ege, Belgium}
\altaffiltext{12}{Charg\' e de recherches, Fonds de la Recherche Scientifique, FNRS, 5 rue d'Egmont, B-1000 Bruxelles, Belgium}

\email{j.j.hermes@warwick.ac.uk}

\begin{abstract}

We present a preliminary analysis of the cool pulsating white dwarf GD\,1212, enabled by more than 11.5\,days of space-based photometry obtained during an engineering test of the two-reaction-wheel-controlled {\em Kepler} spacecraft. We detect at least 19 independent pulsation modes, ranging from $828.2-1220.8$\,s, and at least 17 nonlinear combination frequencies of those independent pulsations. Our longest uninterrupted light curve, 9.0 days in length, evidences coherent difference frequencies at periods inaccessible from the ground, up to 14.5\,hr, the longest-period signals ever detected in a pulsating white dwarf. These results mark some of the first science to come from a two-wheel-controlled {\em Kepler} spacecraft, proving the capability for unprecedented discoveries afforded by extending {\em Kepler} observations to the ecliptic.

\end{abstract}


\section{Introduction}

The endpoints of stellar evolution, white dwarf (WD) stars provide important boundary conditions on the fate of all stars with masses $\leq 8$ \msun, as is the case for more than 97\% of all stars in our Galaxy, including the Sun. When a WD cools to the appropriate effective temperature to foster a hydrogen partial-ionization zone ($\sim12{,}000$\,K, although there is a slight dependence on WD mass), global oscillations driven as non-radial $g$-modes become unstable and reach observable amplitudes (see \citealt{WinKep08,FontBrass08}).

These hydrogen-atmosphere pulsating WDs (so-called DAVs or ZZ\,Ceti stars) have spent hundreds of Myr passively cooling before reaching this evolutionary state. Global oscillations provide a unique window below the thin photosphere and deep into the interior of these stars, enabled by matching the observed periods to theoretical periods generated by adiabatic pulsation calculations.

Given the number of free parameters for full asteroseismic fits, the most reliable results require securing a large number of pulsation periods and uniquely identifying the oscillation modes. However, with typical $g$-mode periods ranging from $100-1400$ s, ground-based photometry suffers from frequent gaps in coverage, frustrating efforts to disentangle multiperiodic signals and alias patterns.

Multi-site campaigns coordinated across the globe via the Whole Earth Telescope (WET, \citealt{Nather90}) have proved the richness of well-resolved WD pulsation spectra. For example, less than a week of nearly continuous observations of the helium-atmosphere (DBV) GD\,358 revealed more than 180 significant periodicities in the power spectrum, providing exquisite constraints on the helium-envelope mass, $(2.0\pm1.0) \times 10^{-6}$\,\mstar, the overall mass, $0.61\pm0.03$\,\msun, and the magnetic field strength, $1300\pm300$\,G \citep{Winget94}. Similarly, multiple WET campaigns on the pre-WD PG\,1159$-$035 have revealed 198 periodicities, accurately constraining the mass, rotation rate and magnetic field of this DOV \citep{Winget91,Costa08}.

\begin{figure*}
\centering{\includegraphics[width=0.985\textwidth]{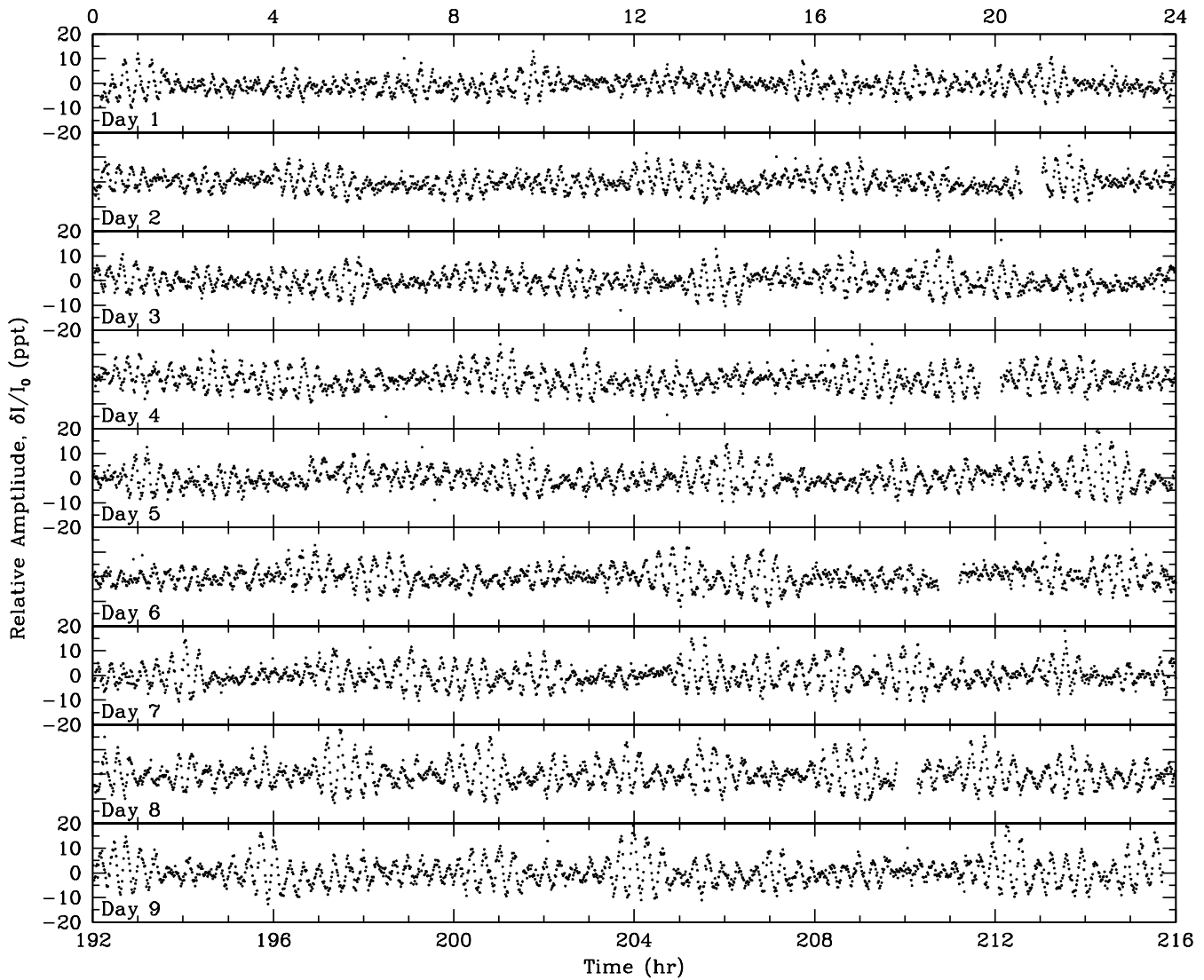}}
\caption{A portion of the minute-cadence photometry collected on the pulsating white dwarf GD\,1212 in 2014~February. This 9.0-day light curve (shown without any smoothing) establishes the capabilities of a two-wheel-controlled {\em Kepler} spacecraft on this $K_p=13.3$ mag white dwarf. An additional 2.6\,days of data, not shown, were collected on GD\,1212 in 2014~January and included in our analysis. \label{fig:GD1212lc}}
\end{figure*}

DAVs have also been extensively studied by more than 10 WET campaigns, with varying results. In part, this is a result of how pulsation modes excited in DAVs are characteristically influenced by the WD effective temperature: hotter DAVs tend to have fewer modes, lower amplitudes and shorter-period pulsations, while cooler DAVs driven by substantially deeper convection zones tend to have more modes at higher amplitude and longer periods (e.g. \citealt{Mukadam06}). WET campaigns have borne this out. More than 5 days of nearly continuous monitoring of the hot DAV G226$-$29 revealed just one significant triplet pulsation \citep{Kepler95}, whereas the cooler DAV G29$-$38 has more than a dozen modes of higher amplitude \citep{Kleinman94}.

In fact, G29$-$38 illustrates the challenges faced to performing asteroseismology of cooler DAVs: although the WD exhibits at least 19 independent oscillation frequencies, there is significant amplitude and phase modulation of these modes, which change dramatically from year-to-year \citep{Winget90,Kleinman98}. Another excellent example of this complex behavior is the cool DAV HL\,Tau\,76, which shows 34 independent periodicities along with many oscillation frequencies at linear combinations of the mode frequencies \citep{Dolez06}. The complex mode amplitude and frequency variations are likely the result of longer-period pulsations having much shorter linear growth times, increasing the prevalence of amplitude and phase changes in cooler DAVs with longer periods (e.g., \citealt{Goldreich99}).

The {\em Kepler} mission has already uniquely contributed to long-term distinctions between the handful of hot and cool DAVs eventually found in the original pointing. The longest-studied by {\em Kepler}, the cool DAV KIC\,4552982 ($10{,}860$\,K, \logg\ $=8.16$) discovered from ground-based photometry \citep{Hermes11}, shows considerable frequency modulation in the long-period modes present between $829-1292$\,s (Bell et al. 2014, in prep.). A much hotter DAV was also observed for six months, KIC\,11911480 ($12{,}160$\,K, \logg\ $=7.94$), which shows at least six independent pulsation modes from $172.9-324.5$\,s that are incredibly stable and evidence consistent splitting from a $3.5\pm0.5$\,day rotation rate \citep{Greiss14}.

After the failure of a second reaction wheel in 2013\,May, the {\em Kepler} spacecraft has demonstrated a mission concept using two-wheel control, observing fields in the direction of the ecliptic. This mission concept aims to obtain uninterrupted observations for approximately 75\,days. As part of an initial test to monitor the two-wheel-controlled pointing behavior on long timescales, short-cadence photometry was collected every minute on the cool DAV GD\,1212 during a preliminary engineering run in 2014\,January and February.

GD\,1212 ($V=13.3$ mag) was discovered to pulsate by \citet{Gianninas06}, with roughly 0.5\% relative amplitude photometric variability dominant at 1160.7\,s. The most recent model atmosphere fits to spectroscopy of GD\,1212 find this WD has a \teff\ $= 10{,}970\pm170$ K and \logg\ $= 8.03\pm0.05$ \citep{Gianninas11}, which corresponds to a mass of $0.62\pm0.03$ \msun\ after incorporating the latest 3D-model corrections of \citet{Tremblay13}. This puts the WD at a distance of roughly 18 pc, although GD\,1212 has the lowest proper motion of any WD within 25 pc of the Sun, $33.6\pm1.0$ mas yr$^{-1}$ \citep{Subsavage09}.

In this paper we provide a preliminary analysis of the unique two-wheel-controlled {\em Kepler} observations of GD\,1212. In Section~2 we outline the observations and reductions. We analyze the independent pulsation modes and nonlinear combination frequencies in Sections~3~and~4, respectively. We conclude in Section~5 with a preliminary asteroseismic interpretation and a discussion of these results in the context of a two-wheel-controlled {\em Kepler} mission observing into the ecliptic.

\section{Observations and Reductions}
\label{sec:observations}

We observed GD\,1212 (GJ\,4355, WD\,2336$-$079, WD\,J233850.74$-$074119.9) for a total of 264.5\,hr using the {\em Kepler} spacecraft in two-wheel mode. With only two working reaction wheels, the spacecraft pointing cannot be stabilized in three axes. By pointing the bore-sight close to the plane of the spacecraft's orbit \citep[within about a degree of the ecliptic,][]{Howell14}, the unconstrained spacecraft roll is placed in equilibrium with respect to the solar pressure, the dominant external force exerted on the spacecraft.

The unconstrained spacecraft roll is in unstable equilibrium, and the pointing needs to be corrected every $3-6$\,hr by firing the thrusters. No data is lost during a thruster firing (known as a reset); the change in attitude causes a minor change in the pointing, which is accounted for by moving the aperture centroid. The spacecraft rolls by no more than 25 arcsec between resets, which corresponds to motion of a star at the extreme edge of the field of view by no more than 1 pixel. The spacecraft can point at a single field for up to 80 days before the angle of the Sun with respect to the solar panels exceeds allowed limits. \citet{Putnam14} provides further details of the capabilities and limitations of two-wheel operation.

We collected data on GD\,1212 from 2014\,Jan\,17 to 2014\,Feb\,13 in short-cadence mode (SC), where each exposure is 58.8\,s. After the first 2.6\,days, the observations were interrupted for 15.1\,days by a safe-mode event and subsequent engineering fault analysis. The final 9.0-day light curve can be found in Figure~\ref{fig:GD1212lc}. We have removed all points falling more than 4$\sigma$ from the light curve mean (112 points), resulting in  2.6- and 9.0-day observations with a duty cycle of 98.2\% and 98.9\%, respectively.

In contrast to the primary mission, where only small masks were placed around each star, the data on GD\,1212 were collected using a $50 \times 50$ pixel ``super aperture.'' We expect the aperture size used in two-wheel mode will decrease as confidence in the spacecraft pointing ability increases. The observed pixels were processed through the {\sc CAL} module of the {\em Kepler} pipeline \citep{Quintana10} to produce target pixel files (TPFs); light curve files were not produced for this engineering data. TPF data for all stars observed during this engineering run are available at the MAST Kepler archive\footnote{\url{http://archive.stsci.edu/missions/k2/tpf\_eng/}}. GD\,1212 is object no. 60017836 in this engineering data.

We estimated a local background for each image using a section of pixels absent of stars and subtracted this. We then created a $50 \times 50$ pixel image summed in time, and fit a 2D Gaussian to this summed image, which defined a standard deviation in the semi-major axis, a standard deviation in the semi-minor axis, and a rotation angle. We held these constant to fix an elliptical aperture but allowed it to move to account for motion of the star. We extracted all pixels within 5 times the standard deviation, weighted by how much of the area of each pixel covered by the ellipse.

Our final 9.0-day dataset is nearly continuous and has a formal frequency resolution of 1.29\,\muhz. The median noise level in the Fourier transform near 500 and 1500 \muhz\ for this 9.0-day run is roughly 0.0036\% (36\,ppm). For the entire 26.7-day (42.8\% duty cycle) dataset on the $K_p=13.3$ mag GD\,1212, the median noise level is roughly 0.0028\% (28\,ppm).

\section{Independent Pulsation Modes}
\label{sec:analysis}

As expected given the relatively cool spectroscopically determined temperature, the pulsation periods excited in GD\,1212 are relatively long, ranging from $828.2-1220.8$\,s. A cool effective temperature is also borne out from model atmosphere fits to the photometry of this WD, which find \teff\ $= 10{,}940\pm320$ K and \logg\ $= 8.25\pm0.03$ \citep{Giammichele12}. 

For our discussion, we will adopt the more precise parameters derived from spectroscopy: \teff\ $= 10{,}970\pm170$ K and \logg\ $= 8.03\pm0.05$. The spectroscopy of GD\,1212 is notable in that much of it was collected in order to identify the then-dominant mode at 1160.7\,s, by measuring the pulsation amplitude as a function of wavelength with time-resolved spectroscopy from the Bok 2.3 m telescope at Kitt Peak National Observatory. However, the identification was inconclusive given the variable, low amplitude of the pulsations \citep{Desgranges08}.

\begin{figure*}
\centering{\includegraphics[width=0.98\textwidth]{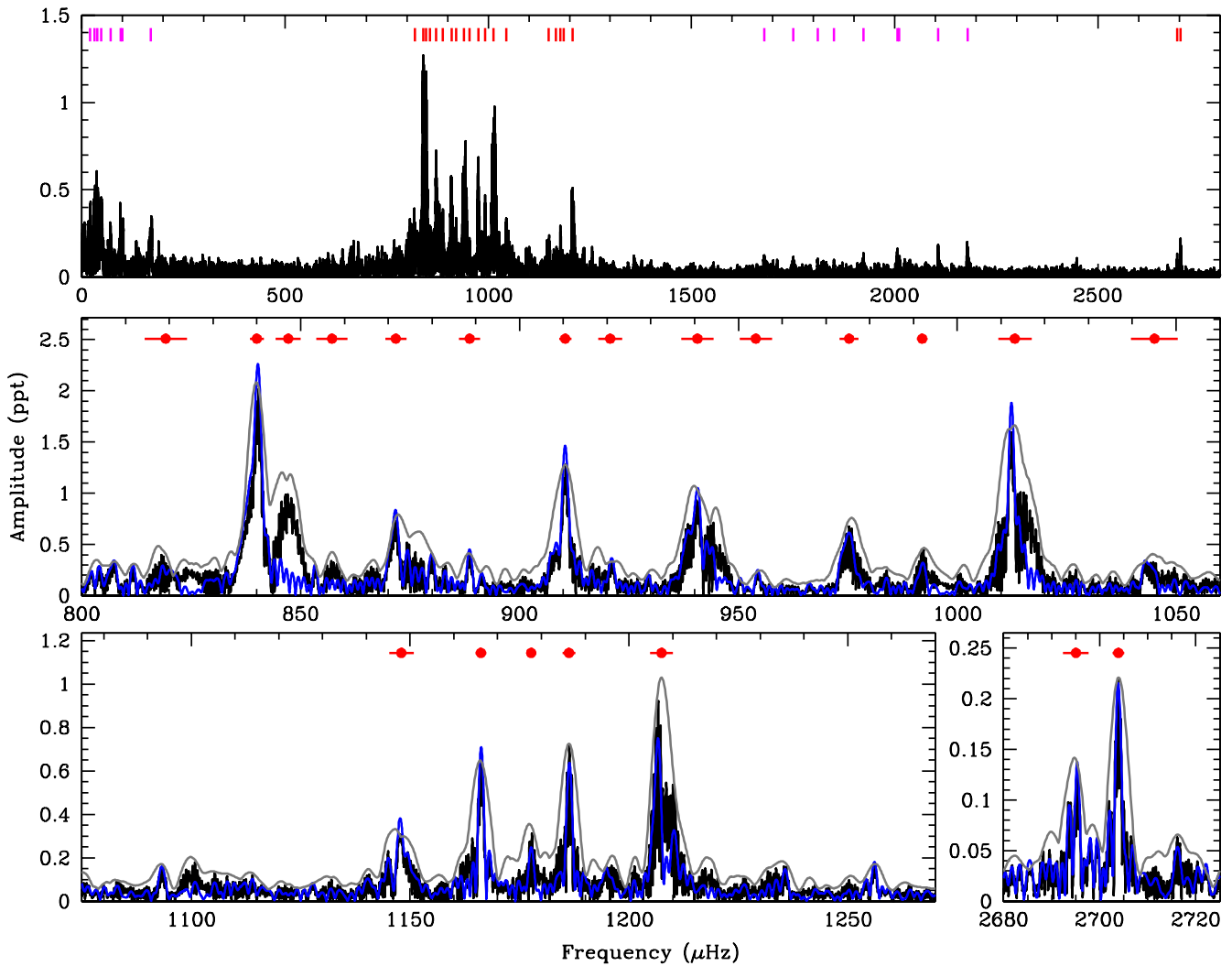}}
\caption{Fourier transforms (FTs) of our two-wheel-controlled {\em Kepler} observations of GD\,1212. The top panel shows the FT of the entire dataset out to 2800 \muhz; there are no significant periodicities at longer frequencies. The middle and bottom panels show the independent pulsation modes in more detail. The blue FT represents only the final 9.0-days of data, while the black FT was calculated from the entire 26.7-day dataset. Additionally, we have split the data into four 2.6-day subsets, and we average the amplitudes of the FTs of each of these subsets, shown in grey for the bottom two panels. We mark the pulsation modes adopted, detailed in Table~\ref{tab:freq}, with red points and associated uncertainties at the top of each panel. Additionally, the magenta lines in the top panel mark nonlinear combination frequencies also detected in our data, shown in more detail in Figure~\ref{fig:GD1212combos}. Amplitudes are expressed in ppt, where 1 ppt = 0.1\% relative amplitude. \label{fig:GD1212ft}}
\end{figure*}

A Fourier transform (FT) of our entire dataset is shown in Figure~\ref{fig:GD1212ft}, which orients us to the dominant frequencies of variability. An FT of our entire dataset finds significant variability ranging from 19.2 \muhz\ (14.5\,hr) down to 2703.9 \muhz\ (369.8\,s), which is shown in full in the top panel of Figure~\ref{fig:GD1212ft}. The only periodicity detected with marginal significance at longer frequencies occurs at 4531.8 \muhz\ (0.12 ppt), which is an instrumental artifact sampling the long-cadence exposures of 29.4\,min that often appears in short-cadence {\em Kepler} data \citep{Gilliland10}. We do not include this signal in our analysis.

The highest-amplitude variability in GD\,1212 occurs in the region between $810-1210$\,\muhz\ ($826-1234$\,s). This region evidences at least 19 independent pulsation modes, which we show in more detail in the bottom panels of Figure~\ref{fig:GD1212ft}. However, the amplitudes of variability in this region are not stable over our 26.7 days of observations, so we additionally display the FT of just our final 9.0-day light curve. We also find a number of nonlinear combination frequencies of these highest-amplitude pulsations, which we discuss further in Section~\ref{sec:nonlinear}.

\begin{figure}
\centering{\includegraphics[width=0.485\textwidth]{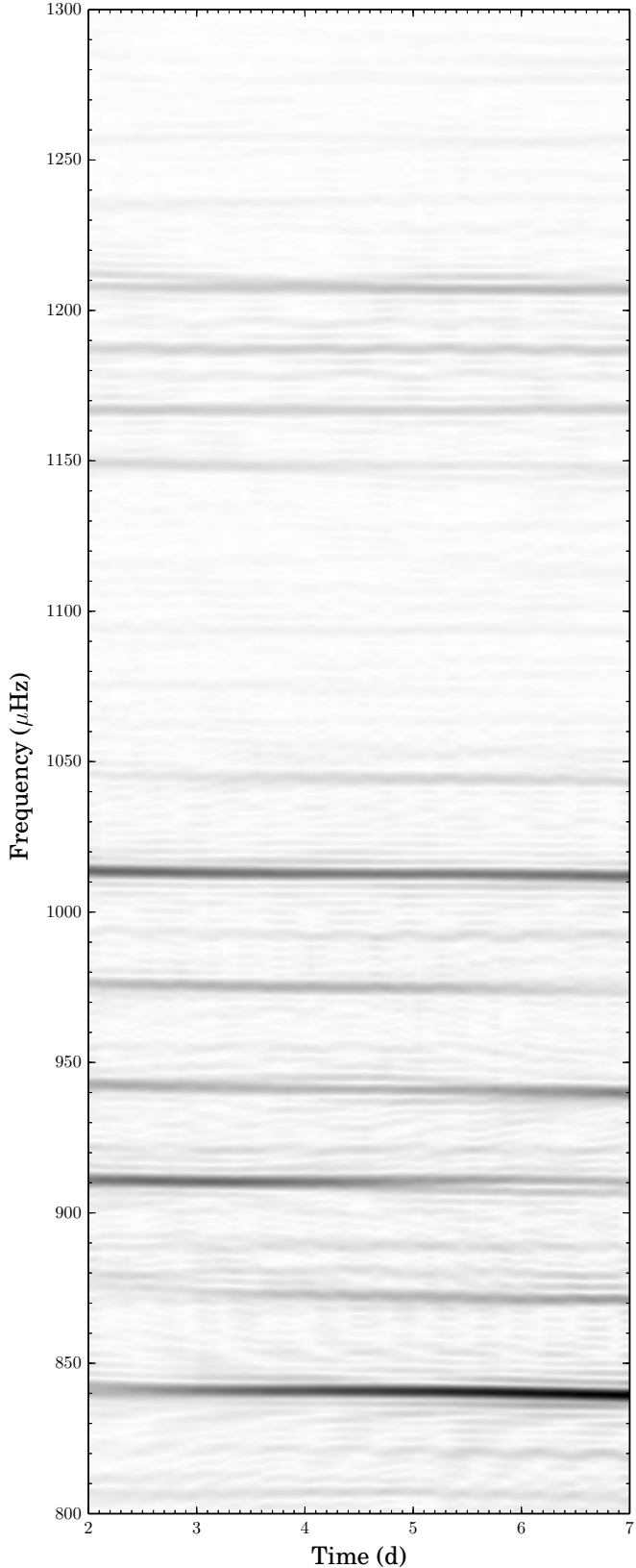}}
\caption{A running FT of the range of independent pulsation modes detected in our 9.0-day, high-duty-cycle two-wheel-controlled {\em Kepler} light curve of GD\,1212. This figure uses a four-day sliding window, and darker greyscale corresponds to higher amplitudes. This demonstrates the lack of amplitude and frequency stability for the pulsations in GD\,1212, even over less than a week. \label{fig:GD1212runningFT}}
\end{figure}

The amplitude and frequency variability of the pulsations in the region between $810-1210$\,\muhz\ are best shown by a running FT of our 9.0-day light curve using a 4.0-day sliding window, shown in Figure~\ref{fig:GD1212runningFT}. For example, the highest-amplitude peak in the FT near 840.0\,\muhz\ is essentially monotonically increasing in amplitude over these last 9.0 days, and is broadly unstable in frequency. Conversely, the running FT shows that the second-highest peak in the FT near 910.5\,\muhz\ is decreasing in amplitude, and appears further to bifurcate.

A common cause of amplitude modulation in pulsating stars is the presence of additional nearby frequencies that are unresolved over the course of observations. The pulsational variability in DAVs is the result of non-radial stellar oscillations, so rotation acts to break the spherical symmetry of a pulsation mode and generates multiplets about the $m=0$ component, spaced by an amount proportional to the rotation rate (e.g., \citealt{Dolez81}). WDs with identified pulsations all vary in $\ell=1,2$ $g$-modes, and thus DAVs which have measured rotational splittings range from $2.5-19.0$ \muhz\ (see \citealt{Kawaler04,FontBrass08} and references therein).

The DAV G226$-$29 is an excellent example of a closely spaced multiplet affecting pulsation amplitudes in a WD \citep{Kepler83}. In this WD, the 109\,s pulsation is actually a superposition of three signals at 109.08648\,s, 109.27929\,s, and 109.47242\,s, which are separated by roughly 16.1\,\muhz\ and require more than 17.2\,hr of observations to resolve the beat period of the closely spaced signals. Shorter observations show the signal sinusoidally varying at the beat period of 17.2\,hr.

However, rotational splittings are unlikely to explain the incoherent changes in the running FT of GD\,1212, nor the disappearance of $f_5$ near 847.2 \muhz, which was strong during the first 2.6 days and was virtually unseen in our final 9.0 days of data. Instead, we are likely observing genuine amplitude and frequency variability internal to the star, as has been witnessed during long campaigns on other cool DAVs (e.g., G29$-$38, \citealt{Kleinman98}) and DBVs (e.g., GD358, \citealt{Provencal09}).

The frequency, amplitude, and possibly phase variability observed in GD\,1212 cause difficulty in defining the true pulsation periods. For hotter DAVs, which show exceptionally stable oscillations, pre-whitening the light curve by the highest peaks in an FT removes most power in that region and allows for a relatively simple extraction of the periods present in the star (e.g., \citealt{Greiss14}). However, none of the peaks in either the 26.7-day FT or the 9.0-day FT of GD\,1212 can be smoothly pre-whitened in the standard way, because fitting a pure sine wave does not completely remove the pulse shape.

One can always decompose the broad peaks in an FT with a linear combination of sine waves, but not all of these sine waves represent physical modes. For example, using just the final 9.0 days of data, the highest-amplitude signal near 840.0\,\muhz\ requires five sine waves to reproduce the broad peak: 840.21 \muhz\ (2.48 ppt), 839.18 \muhz\ (1.54 ppt), 842.02 \muhz\ (1.10 ppt), 838.00 \muhz\ (0.82 ppt), and 836.83 \muhz\ (0.38 ppt). This series of pre-whitened peaks in all likelihood corresponds to only one independent pulsation mode.

\begin{deluxetable}{lccc}
\tablecolumns{4}
\tablewidth{0.45\textwidth}
\tablecaption{Frequency solution for GD 1212
  \label{tab:freq}}
\tablehead{\colhead{ID} & \colhead{Period} & \colhead{Frequency} & \colhead{Amplitude} 
\\ \colhead{} & \colhead{(s)} & \colhead{($\mu$Hz)} & \colhead{(ppt)} }
\startdata
\multicolumn{4}{c}{\bf Independent Pulsation Modes} \\
$f_{14}$ & 1220.75 $\pm$ 7.15 & 819.17 $\pm$ 4.80 & 0.23 \\
$f_{1}$ & 1190.53 $\pm$ 2.28 & 839.96 $\pm$ 1.61 & 2.11 \\
$f_{4}$ & 1180.40 $\pm$ 4.02 & 847.17 $\pm$ 2.89 & 0.85 \\
$f_{18}$ & 1166.67 $\pm$ 4.81 & 857.14 $\pm$ 3.54 & 0.21 \\
$f_{7}$ & 1147.12 $\pm$ 3.19 & 871.75 $\pm$ 2.42 & 0.52 \\
$f_{17}$ & 1125.37 $\pm$ 3.01 & 888.60 $\pm$ 2.38 & 0.22 \\
$f_{2}$ & 1098.36 $\pm$ 1.65 & 910.45 $\pm$ 1.37 & 1.04 \\
$f_{12}$ & 1086.12 $\pm$ 3.27 & 920.71 $\pm$ 2.78 & 0.24 \\
$f_{5}$ & 1063.08 $\pm$ 4.13 & 940.66 $\pm$ 3.66 & 0.63 \\
$f_{20}$ & 1048.19 $\pm$ 4.01 & 954.02 $\pm$ 3.65 & 0.13 \\
$f_{9}$ & 1025.31 $\pm$ 2.26 & 975.31 $\pm$ 2.15 & 0.49 \\
$f_{11}$ & 1008.07 $\pm$ 1.20 & 991.99 $\pm$ 1.18 & 0.31 \\
$f_{3}$ & 987.00 $\pm$ 3.73 & 1013.17 $\pm$ 3.82 & 0.95 \\
$f_{16}$ & 956.87 $\pm$ 4.91 & 1045.08 $\pm$ 5.36 & 0.22 \\
$f_{15}$ & 871.06 $\pm$ 2.13 & 1148.02 $\pm$ 2.81 & 0.22 \\
$f_{8}$ & 857.51 $\pm$ 0.86 & 1166.17 $\pm$ 1.17 & 0.49 \\
$f_{13}$ & 849.13 $\pm$ 0.76 & 1177.68 $\pm$ 1.05 & 0.23 \\
$f_{10}$ & 842.96 $\pm$ 1.02 & 1186.30 $\pm$ 1.43 & 0.46 \\
$f_{6}$ & 828.19 $\pm$ 1.79 & 1207.45 $\pm$ 2.61 & 0.53 \\
$f_{21}$ & {\em 371.05 $\pm$ 0.36} & {\em 2695.04 $\pm$ 2.63} & {\em 0.07} \\
$f_{19}$ & {\em 369.83 $\pm$ 0.17} & {\em 2703.91 $\pm$ 1.24} & {\em 0.15} \\
\multicolumn{4}{c}{\bf Nonlinear Combination Frequencies} \\
$f_{10}-f_8$ & $50{,}500 \pm 10{,}600$ & 19.82 $\pm$ 4.16 & 0.24 \\
$f_7-f_1$ & $31{,}540 \pm 3580$ & 31.70 $\pm$ 3.60 & 0.27 \\
$f_2-f_7$ & $25{,}990 \pm$ 2870 & 38.47 $\pm$ 4.24 & 0.35 \\
$f_{18}-f_1$ & $20{,}920 \pm$ 2280 & 47.80 $\pm$ 5.21 & 0.30 \\
$f_2-f_1$ & $14{,}210 \pm$ 539 & 70.37 $\pm$ 2.67 & 0.21 \\
$f_5-f_1$ & $10{,}463 \pm$ 428 & 95.57 $\pm$ 3.91 & 0.21 \\
$f_3-f_2$ & 9967 $\pm$ 274 & 100.33 $\pm$ 2.75 & 0.25 \\
$f_3-f_1$ & 5910 $\pm$ 162 & 169.19 $\pm$ 4.66 & 0.22 \\
$2f_1$ & 595.45 $\pm$ 1.29 & 1679.41 $\pm$ 3.65 & 0.08 \\
$f_1+f_2$ & 571.18 $\pm$ 1.30 & 1750.75 $\pm$ 4.00 & 0.08 \\
$f_5+f_7$ & 552.36 $\pm$ 0.68 & 1810.41 $\pm$ 2.23 & 0.08 \\
$f_1+f_3$ & 540.23 $\pm$ 0.64 & 1851.07 $\pm$ 2.18 & 0.09 \\
$f_2+f_3$ & 519.94 $\pm$ 0.88 & 1923.29 $\pm$ 3.27 & 0.09 \\
$f_1+f_8$ & 498.36 $\pm$ 0.88 & 2006.60 $\pm$ 3.54 & 0.11 \\
$f_4+f_8$ & 497.13 $\pm$ 0.99 & 2011.53 $\pm$ 4.01 & 0.08 \\
$f_5+f_8$ & 474.56 $\pm$ 0.45 & 2107.22 $\pm$ 1.99 & 0.12 \\
$f_3+f_8$ & 458.89 $\pm$ 0.49 & 2179.19 $\pm$ 2.35 & 0.12
\enddata
\end{deluxetable}

\begin{figure*}
\centering{\includegraphics[width=0.98\textwidth]{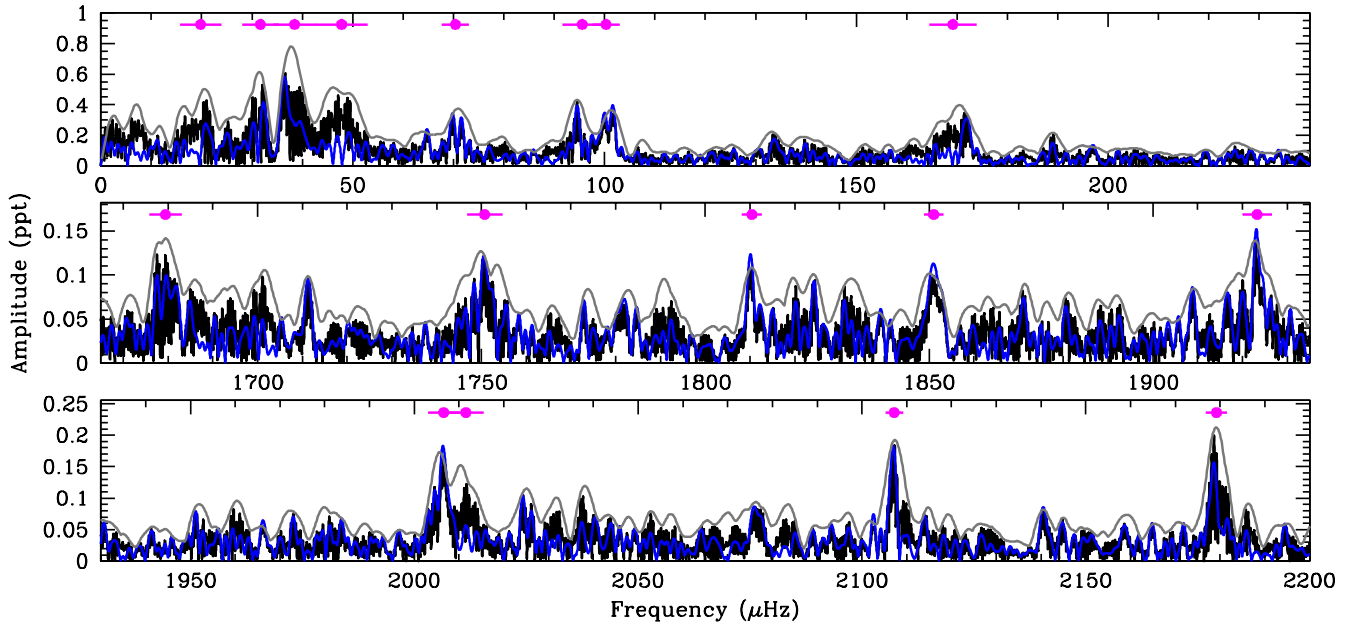}}
\caption{Fourier transforms (FTs) of the nonlinear combination frequencies detected in our observations of GD\,1212. As in Figure~\ref{fig:GD1212ft}, the blue FT represents only the final 9.0-days of data, while the black FT was calculated from the entire 11.6 days of data spread over 26.7 days. Additionally, the average FT of four 2.6-day subsets is shown in grey. We mark the adopted combination frequencies, detailed in Table~\ref{tab:freq}, with magenta points and associated uncertainties at the top of each panel. \label{fig:GD1212combos}}
\end{figure*}

The pre-whitening method does provide a quantitative test for significance. We can iteratively fit and pre-whiten the highest peaks in the FT until there are none above some threshold; we estimate this locally for each frequency by calculating the median value, $\sigma$, of the FT in a $\pm200$\,\muhz\ sliding window, and mark as significant the signals that exceed 5$\sigma$. Since the 15.1-day gap in our whole dataset strongly degrades the window function, we have adopted this approach for our final 9.0 days of data and our initial 2.6-day light curve. This provides an input list of 48 significant signals.

We do not, however, use the pre-whitening method to define adopted periods and the associated period uncertainties of the independent pulsation modes, which form the input for asteroseismic modeling. Performing a linear least-squares fit of the periods determined from the iterative pre-whitening technique would greatly underestimate the period uncertainties.

Instead, we have fit a Lorentzian profile to a $\pm5$\,\muhz\ range of the regions of power defined significant using our pre-whitening method, and use the central peak, half-width-half-maximum, and intensity of this Lorentzian to define the adopted periods, period uncertainties, and amplitudes of the broad peaks in the FT. These values provide the red points and uncertainties shown in Figure~\ref{fig:GD1212ft} that conservatively represent the periods of excited variability, and are detailed in Table~\ref{tab:freq}. Lorentzian profiles are frequently used to describe stochastic oscillations driven by turbulent convection in cool main-sequence and red giant stars, but our use of these profiles is primarily out of convenience and does not suggest that stochastic driving is undergoing in GD\,1212 (a discussion of driving in DAVs can be found in \citealt{Brickhill91}).

This method strongly underestimates the adopted pulsation amplitudes, but theoretical pulsation calculations do not account for mode amplitudes, and our results reasonably represent the two quantities requisite for a seismic analysis: the pulsation periods and associated uncertainties. We italicize the signals at 371.1\,s and 369.8\,s in Table~\ref{tab:freq} because they may not in fact be independent pulsation modes (see discussion at the end of Section\,\ref{sec:nonlinear}).

When establishing the coherence of variability over long timescales, it is often instructive to calculate FTs from multiple different subgroups of the data of identical resolution and then average the amplitudes of the FTs of each of these subsets. This effectively throws away the phase information of the different datasets, and identifies the most coherent regions of the FT. Figure~\ref{fig:GD1212ft} shows such an average FT for four equal-length (2.6-day) subsets of our data on GD\,1212. We find that smaller 2.6-day subsets are not sufficient to resolve some closely spaced variability that can be resolved within longer subsets, such as the FT of our 9.0-day light curve.

It is possible that this frequency broadening for the averaged FTs is the result of unresolved rotational multiplets embedded within the natural frequency variability of the underlying pulsation modes. The average half-width-half-maximum of the profiles fit to the highest seven peaks of the averaged FT is 3.14\,\muhz, compared to 1.12\,\muhz\ for the 9.0-day FT. This difference could be explained by rotational splittings arising from a WD rotation rate between $1.9-5.2$ days if these are $\ell=1$ modes and between $6.1-17.2$ days if these are $\ell=2$ modes.

An inferred rotation rate of $1.9-17.2$ days provides one of the first constraints on the rotation rate of a cool DAV, albeit a very coarse estimate. Hotter, stable DAVs show detected rotation rates from rotational splittings between $0.4-2.3$\,days \citep{Kawaler04,FontBrass08}.

\section{Nonlinear Combination Frequencies}
\label{sec:nonlinear}

In addition to the 19 significant independent pulsation modes we identify in Section~\ref{sec:analysis}, we see a number of nonlinear combination frequencies in our data that arise at summed and difference frequencies of the independent modes. For example, the peak at 2107.3\,\muhz\ is, within the uncertainties, a linear sum of the two independent pulsations $f_4=940.3$\,\muhz\ and $f_7=1166.2$\,\muhz.

Combination frequencies are a common feature in the pulsation spectra of variable WDs, which are created by a nonlinear distortion of an underlying linear oscillation signal. Work by \citet{Brickhill92} attributed these nonlinear distortions to the changing thickness of the convection zone of a DAV, a result of local surface temperature variations from the underlying global stellar oscillations. The convection zone depth is extremely sensitive to temperature ($\propto T^{-90}$, \citealt{Montgomery05}), distorting the observed emergent flux. For the hottest DAVs, the nonlinear response of the flux to a temperature perturbation (i.e., the ``$T^4$'' nonlinearity) may also play a role in creating these signals \citep{Brassard95}.

Observations of the rate of period change of combination frequencies show that they match identically the rates of their parent modes, establishing that these signals are not independently excited pulsations but rather nonlinear distortions \citep{Hermes13}. Thus, these signals provide no additional asteroseismic constraints on the interior structure of the star. However, the amplitude ratios of these combination frequencies to their parent modes can yield insight into the depth of the convection zone, which is responsible for driving the underlying independent pulsations (e.g., \citealt{Montgomery10}).

We detail the observed nonlinear combination frequencies in GD\,1212 and identify their likely parent modes in Table~\ref{tab:freq}. Their amplitudes and frequencies were determined in an identical manner to the independent pulsation modes, as discussed in Section~\ref{sec:analysis}. However, we have used a less stringent test for significance, and include as significant the signals that exceed 4$\sigma$. We mark these adopted nonlinear combination frequencies as magenta points in our FT of all the data, shown in the top panel of Figure~\ref{fig:GD1212ft}. We show a more detailed view of the regions of these combination frequencies in Figure~\ref{fig:GD1212combos}.

Good evidence that these are in fact combination frequencies comes qualitatively from the region in the FT near 2006\,\muhz\ (bottom panel of Figure~\ref{fig:GD1212combos}) as compared to the region near 847\,\muhz\ (middle panel, Figure~\ref{fig:GD1212ft}). There appear two combination frequencies, 2006.6\,\muhz\ and 2011.5\,\muhz, which are combination of $f_1+f_7$ and $f_5+f_7$, respectively. The power for the $f_5+f_7$ combination frequency is nearly nonexistent in our final 9.0 days of data, exactly as the power for $f_5$ near 847.2\,\muhz\ is diminished in our final 9.0 days of data.

Unfortunately, the underlying amplitude and frequency variability of the independent parent modes extends to these combination frequencies, broadening their power in the FTs. Therefore, these nonlinearities do not provide much assistance in refining the periods of the parent modes. Still, the eight significant low-frequency difference frequencies, ranging from $19.2-170.0$\,\muhz\ ($1.6-14.5$ hr), are the longest-period signals ever detected in a pulsating WD, accessible only because of the exceptionally long and uninterrupted space-based observations provided by the {\em Kepler} spacecraft.

We preliminarily identify the variability at $f_{21}=371.1$\,s and $f_{19}=369.8$\,s as independent pulsation modes, but they arise at substantially shorter periods, far isolated from the other independent pulsations, which are all longer than 828.2\,s. The signals $f_{21}$ and $f_{19}$ may instead be more complicated nonlinear combination frequencies. For example, $f_{21}$ may be the sum $2f_1+f_2$. However, no obvious frequency combination can produce the higher-amplitude $f_{19}$.

It is also possible that the signal near 370\,s may be a new two-wheel-controlled {\em Kepler} instrumental artifact. However, we do not observe a significant peak near 2700\,\muhz\ in any other short-cadence targets observed during the two-wheel-controlled {\em Kepler} engineering test run.

\section{Discussion and Conclusions}
\label{sec:discussion}

We report a detailed census of pulsation modes excited in the cool DAV GD\,1212 using more than 11.5 days of data collected during a two-wheel concept engineering test of the repurposed {\em Kepler} spacecraft. Our results suggest the presence of at least 19 independent pulsation modes excited to observable amplitudes, with up to 2.5\% peak-to-peak variability.

Observed frequency and amplitude modulation complicates a precise determination of the pulsation periods in GD\,1212, likely the result of these relatively long-period oscillations ($828.2-1220.8$\,s) having short linear growth times. We have defined a conservative estimate of the excited pulsations by fitting Lorentzian functions to the broad bands of power in an FT of our entire dataset, yielding an average uncertainty of 3.0\,s ($<0.5$\%) for each adopted period listed in Table~\ref{tab:freq}.

A complete asteroseismic solution of the data in hand would be premature until we have more secure identification of the spherical degree ($\ell$ values) of the independent pulsation modes, the subject of a future paper. We can make a first attempt at these identifications given that such long-period modes likely have high radial order ($k>17$). This is near the asymptotic limit, in which patterns in consecutive period spacing can yield insight into the spherical degree of the modes.

There appear to be two regions of semi-evenly spaced independent pulsations, one at longer periods (involving $f_1$-$f_7$-$f_2$-$f_5$-$f_9$-$f_3$) that have a mean period spacing of $41.5\pm
2.5$\,s. Additionally, four shorter-period modes ($f_{15}$-$f_8$-$f_{10}$-$f_6$) have a spacing of $14.4\pm1.5$\,s. The asymptotic period spacing for $\ell=1$ modes of a cool ($11{,}147$\,K), 0.837\,\msun WD with a  $10^{-5.408}$\,\mstar\ hydrogen layer mass is roughly 40.3\,s, according to the models of \citet{Romero12}. However, this model does not accurately predict the other period spacings, as they calculate $\ell=2$ modes to have a 23.3\,s mean period spacing.

The asymptotic period spacing for higher spherical degrees decreases as $\sqrt{\ell(\ell+1)}$, so it is possible that the shorter-period sequence is composed of a series of $\ell=3$ modes. However, $\ell=3$ modes have never been previously inferred in WDs, although $\ell=4$ modes have been proposed in G185-32 \citep{Thompson04}. Limb-darkening and geometric effects greatly reduce the observed amplitudes of $\ell=1,2$ modes compared to $\ell=3$ modes at shorter (near-UV) wavelengths (e.g., \citealt{Kepler00}), so multicolor observations of GD\,1212 could conclusively test this hypothesis.

It is perhaps more likely that the two seemingly evenly spaced regions are not consecutive radial order modes of the same spherical degree at all, but rather a coincidence of overlapping $\ell=1$ and $\ell=2$ modes. A coarse grid search using the models described in \citet{Castanheira08}, allowing the highest-amplitude periodicities to be $\ell=1,2$, can modestly reproduce the observed periods if $f_8,f_3,f_5,f_7,f_1$ are $\ell=1$ modes and if $f_6,f_{10},f_{15},f_2$ are $\ell=2$ modes, if the WD is $10{,}700$ K, 0.81 \msun, and has a $10^{-5.5}$ \mstar\ hydrogen layer.

Another interpretation of the period-spacing patterns suggests that all modes with periods in excess of 956\,s are $\ell=1$ modes, with overtones spaced by about 37\,s. Each overtone would appear as an $m = \pm 1$ doublet split in total by roughly 16.5\,\muhz. Modes with shorter periods roughly follow a series spaced by about 21\,s (consistent with $l=2$ given the $l=1$ spacing for the
longer-period modes), with each overtone expressed as a doublet with roughly 19.6\,\muhz\ splittings. It is not uncommon to observe only the $\pm1$ components of an $l=1$ triplet and not the $m=0$ central component, due to inclination or excitation effects (e.g. \citealt{Winget91}).

More quantitative answers will require a full asteroseismic analysis, matching all 19 independent pulsation modes to the periods expected from theoretical models, which should explicitly reveal the overall mass, temperature, and hydrogen-layer mass of this cool pulsating WD.

As DAVs cool and approach the red edge of the ZZ Ceti instability strip, the observed pulsation periods generally increase, often exhibiting large-amplitude, nonlinear pulse shapes \citep{Kanaan02}. The periods observed in the cool ($10{,}970$ K) GD\,1212 fit this trend: the weighted mean period of the observed pulsations is roughly 1061\,s, one of the longest observed for any DAV \citep{Mukadam06}. However, the pulsation amplitudes and nonlinearities observed in GD\,1212 are notably weak, suggesting this WD is a somewhat rare low-amplitude DAV near the red edge.

Our final 9.0-day light curve displayed in Figure~\ref{fig:GD1212lc} is, to date, the highest-quality continuous look at a pulsating WD, unmolested by differential atmospheric effects, clouds, or sunrises. Much of the long-term variability about the mean of the light curve is real, caused by nonlinear combination frequencies of the excited pulsation modes at periods extending up to 14.5\,hr. Such long-period signals have never been detected from ground-based campaigns.

This relatively short engineering test proves the impact that a two-wheel-controlled {\em Kepler} spacecraft, observing in the ecliptic, can have in making new discoveries in the exploration of stellar interiors and asteroseismology. There are at least 36 known pulsating WDs brighter than $g<18.0$ mag within 12 deg of the first nine proposed {\em K2} fields (see \citealt{Howell14}). We expect that such a mission will afford us unprecedented insight into the heretofore elusive cause of why pulsations in DAVs shut down at the red edge, as well as uncovering additional low-amplitude pulsations simply inaccessible from ground-based observations that will further constrain asteroseismic modeling of these degenerate stars.

\acknowledgments

We acknowledge Working Group 11 of the {\em Kepler} Asteroseismic Science Consortium (KASC) for the eager discussion and analysis that led to this rapid result. We thank the referee, S.O. Kepler, for helpful comments that improved aspects of the discussion. J.J.H. acknowledges funding from the European Research Council under the European Union's Seventh Framework Programme (FP/2007-2013) / ERC Grant Agreement n. 320964 (WDTracer). M.H.M. and D.E.W. gratefully acknowledge the support of the NSF under grants AST-0909107 and AST-1312678 and the Norman Hackerman Advanced Research Program under grant 003658-0252-2009. M.H.M. acknowledges the support of NASA under grant NNX12AC96G, and D.E.W. acknowledges the support of NASA under grant NNX13AC23G. S.B. is supported by the Foundation for Fundamental Research on Matter (FOM), which is part of the Netherlands Organisation for Scientific Research (NWO). D.H. acknowledges support by NASA under grant NNX14AB92G issued through the {\em Kepler} Participating Scientist Program.
	
{\it Facilities:} Kepler, K2

\end{document}